           \documentstyle[prl,preprint,aps]{revtex}

\begin{document}

\title{
                     Astrophysical Solutions are Incompatible \\
                     with the Solar Neutrino Data
}
\author{
                     S.\ Bludman, N.\ Hata, and P. Langacker
}
\address{                                                         
                     Department of Physics,                       
                     University of Pennsylvania,                  
                     Philadelphia, PA 19104                       
}                                                                 
\date{
                     June 1, 1993, UPR-0572T
}
\maketitle


\begin{abstract}

We consider the most general solar model, using the neutrino fluxes as free
parameters constrained only by the solar luminosity, and show that the
combined solar neutrino data exclude any astrophysical solution at 98\% C.L.\
Our best fit to the $^7$Be and $^8$B fluxes is respectively $<$7\%  and
37$\pm$4\% of the standard solar model prediction, but only with a large
$\chi^2$ (5.6 for 1 d.f.). This best fit to the fluxes contradicts explicit
nonstandard solar models, which generally reduce the  $^8$B flux more than the
$^7$Be.  Those models are well parameterized by a single parameter, the
central temperature.

\end{abstract}

\newpage


Each of the solar neutrino experiments \cite{Homestake,KII,KIII,SAGE,GALLEX}
show deficits of the solar neutrino flux
compared to the standard solar model (SSM) predictions
\cite{Bahcall-Pinsonneault,Turck-Chieze-Lopes} as summarized in Table~I.
Numerous astrophysical solutions have been proposed to explain the
discrepancy between theory and experiments
\cite{Bahcall-Ulrich,Dearborn,Castelliani,sixteen,WIMPs}.
One category of such proposals changes the input
parameters of the solar models, assuming that the uncertainties of those
quantities might be significantly underestimated in the SSMs.
For example, the neutrino fluxes are known to be
sensitive to the opacity, and explicit models have been constructed with
smaller values of the opacity or smaller values of the heavy element
abundance.  The nuclear reaction
cross sections,  which are extrapolated from laboratory conditions, are
another potential source of uncertainties: there might be some mechanism
affecting the low energy cross sections and reducing the neutrino
production, and such effects might even be  correlated among
the different reactions.  A second  category of proposals attributes
the neutrino deficit to  mechanisms
such as rotation, magnetic fields,
turbulent diffusion, mixing of elements, or hypothetical weakly
interacting particles (WIMPs) that are not included in the SSMs.
Both kinds of theoretical proposals usually reduce the expected flux of
high- and medium-energy neutrino production by
lowering
the temperature in the core region where the nuclear fusion takes place, and
can be parameterized by a lower central temperature ($T_C$).
In previous studies using the power law dependence of the neutrino fluxes on
$T_C$ \cite{BKL,BHKL,GALLEX},  it was shown quantitatively that such cooler
sun models are incompatible with the experimental data, especially because
the higher
Kamiokande observed rate relative to the Homestake rate cannot be explained
so long as cool sun models reduce the expected $^8$B flux more
than the $^7$Be flux.
  This failure of astrophysical resolutions of the solar neutrino deficit
suggests particle physics solutions, such as
the Mikheyev-Smirnov-Wolfenstein (MSW) mechanism \cite{MSW}, which fits
all observations and is taken as a strong hint of neutrino mass and mixings
\cite{BHKL,HL,Shi-Schramm,Gelb-Kwong-Rosen,Krastev-Petcov,Krauss-Gates-White}.

In this paper, we remove the assumption of a power law dependence
and examine arbitrary solar models by allowing the four relevant
neutrino fluxes $\phi(pp)$, $\phi(\mbox{Be})$, $\phi(\mbox{B})$, and
$\phi(\mbox{CNO})$ to change freely.  We do not advocate
such models or claim that they are consistent with  other solar observations,
but only show that they are incompatible with the solar neutrino data.
In our  most general solar model, we
assume: (1) The Sun is in quasi-static equilibrium and generates energy
by nuclear fusion in the $pp$  and the CNO chains;  (2) Astrophysical
mechanisms may change the magnitude of each neutrino flux component, but do
not significantly distort the energy spectra of the individual components
\cite{Bahcall-spectrum}.
(Particle physics solutions such as the MSW effect often depend
on  neutrino energy and therefore do distort the neutrino spectrum.);
(3) All reported experimental results are correct,  as well as the
calculations of radio-chemical detector cross sections.  Because  the
Kamiokande and Homestake results are crucial to  our conclusions,  we will
also consider the possibilities that their uncertainties have been
underestimated.

By our first assumption, the well-measured solar luminosity imposes the
constraint
\begin{equation}
\label{L_const}
     \phi(pp) + 0.979 \, \phi(\mbox{Be}) + 0.955 \, \phi(\mbox{CNO})
   = 6.57 \times 10^{10} \, \mbox{cm}^{-2}\mbox{s}^{-1},
\end{equation}
among the $pp$, $^7$Be, and CNO fluxes, when the different energies carried
off by neutrinos  are taken into account.

In Figs.~\ref{fig_exps}--\ref{fig_xsec} we present  the results of all
solar neutrino experiments in the
$\phi(\mbox{Be})-\phi(\mbox{B})$
plane in units of the Bahcall-Pinsonneault predicted fluxes.
Essentially all astrophysical solutions, including insensible models,
are represented in the plane, from
$\phi(\mbox{Be})/\phi(\mbox{Be})_{SSM}= \phi(\mbox{B})/\phi(\mbox{B})_{SSM}=1$
for the SSM to
$\phi(\mbox{Be})/\phi(\mbox{Be})_{SSM}= \phi(\mbox{B})/\phi(\mbox{B})_{SSM}=0$
for the minimum rate model \cite{minimum-rate-model}.
Fig.~\ref{fig_exps} shows the constraints from each experiment obtained by
minimizing the $\chi^2$
with respect to $\phi(pp)$ and $\phi$(CNO) at each point subject to the
luminosity constraint.  Our $\chi^2$ fit includes experimental uncertainties
as well
as detector cross section uncertainties.  The uncertainties of minor fluxes
($pep$, $hep$, and $^{17}$F) are included, but contribute negligibly.
The Kamiokande result constrains only the $^8$B
flux, while the Homestake data and combined SAGE-GALLEX  data  constrain both
the $^7$Be and $^8$B fluxes.  At 90\% confidence level (C.L.), none of the
experiments are consistent with the Bahcall-Pinsonneault SSM. Indeed, the
combined experiments together allow only a small parameter space around
$\phi(\mbox{Be})/\phi(\mbox{Be})_{SSM} \sim 0$
and
$\phi(\mbox{B})/\phi(\mbox{B})_{SSM} \sim 0.4$,
but with a large $\chi^2$ value: 5.6 for 0 degrees of freedom
(3 data -- (4 parameters -- 1 constraint)).  No general statistical
interpretation exists in such a case, other than to conclude that
this model is excluded.
If one considers the $^7$Be flux to be fixed at 0, then the fit has 1 degree
of freedom and this possibility is excluded at the 98\% C.L.\ This
shows that any astrophysical solution in which the spectral
shape of the individual neutrino fluxes is unchanged is incompatible with
observations.

Fig.~\ref{fig_comb}  shows the confidence levels of the combined fit
in the two dimensional $\phi(\mbox{Be})-\phi(\mbox{B})$ subspace.
The contours are determined by $\chi^2 = \chi^2_{min} + \Delta\chi^2$, where
$\chi^2_{min}$ is obtained allowing an unphysical negative $^7$Be flux.
We could alternately have taken $\chi^2_{min}$ in the physical region
by restricting the probability distribution to the
physical region ($\phi(\mbox{Be}) \geq 0$).   This procedure would have ignored
the fact that the best fit is very poor, and would grossly overestimates the
allowed region.  We therefore present our results as a qualitative display
of the confidence levels.

We  also show in Fig.~\ref{fig_comb} the Bahcall-Pinsonneault SSM with
90\% C.L.
uncertainties, the 1000 Monte-Carlo SSMs of Bahcall and Ulrich
\cite{Bahcall-Ulrich}, the central value of the Turck-Chi\`eze-Lopes (TC) SSM
\cite{Turck-Chieze-Lopes}, and various explicit nonstandard solar models
constructed to solve the solar neutrino problem: the low Z model in which
the heavy element abundance is reduced by 90\% from the standard value
\cite{Bahcall-Ulrich};
the low opacity models with 10 and 20\% reduced opacity \cite{Dearborn};
the solar models with increased $pp$
cross sections ($S_{11}$) by 30, 50, 80, 100, and 150\% from the SSM value
\cite{Castelliani}; and the solar model with WIMPs
\cite{WIMPs,Bahcall-Ulrich}.
Also the power laws for the core temperature and $S_{11}$ obtained
from the Monte-Carlo SSMs are extrapolated from the SSM region and displayed.
The uncertainty due to the $p$+$^7$Be cross section (9.3\%) is shown as
error bars.

Because the decay of $^8$B follows the reaction
$p + ^7\mbox{Be} \rightarrow ^8\mbox{B} + \gamma$,
all explicit nonstandard models predict more reduction of
the $^8$B flux than the $^7$Be flux ({i.e.,}
$\phi(\mbox{Be})/\phi(\mbox{Be})_{SSM} > \phi(\mbox{B})/\phi(\mbox{B})_{SSM}$).
Any reduction of the $^7$Be
production rate affects both the $^8$B and $^7$Be flux equally.
Other uncertainties in the $p$+$^7$Be rate affect only the $^8$B flux.
Therefore, unless there is
some independent mechanism to suppress only the $^7$Be neutrino emission,
all realistic nonstandard solar models are in serious contradiction to the
solar neutrino data, which constrain the two fluxes to
$\phi(\mbox{Be})/\phi(\mbox{Be})_{SSM} < 0.07$
and
$\phi(\mbox{B})/\phi(\mbox{B})_{SSM} = 0.37 \pm 0.04$ (1$\sigma$).
This emphasizes that there are {\it two} solar neutrino problems:
(I) The neutrino fluxes observed in every experiment are significantly below
SSM predictions at 90\% C.L. (II) Kamiokande and Homestake together allow only
the very implausible fit
$\phi(\mbox{Be})/\phi(\mbox{Be})_{SSM} \ll
 \phi(\mbox{B})/\phi(\mbox{B})_{SSM}$.

The two curves in Figs.~\ref{fig_comb}--\ref{fig_xsec}
assume that the $^7$Be and $^8$B neutrino fluxes each depend simply on
powers of the central temperature $T_c$ (solid curve) or of the $pp$ nuclear
cross section factor $S_{11}$ (dot-dashed curve), while the $pp$ and CNO
neutrino fluxes are adjusted to obey the solar luminosity  constraint
(Eq.~\ref{L_const}).  For the solid curve we assumed
$\phi(\mbox{Be}) \sim T_c^8, \phi(\mbox{B}) \sim T_C^{18}$
so that
$\phi(\mbox{B}) \propto \phi(\mbox{Be})^{2.25}$.
For the dot-dashed curve  we assumed
$\phi(\mbox{Be}) \sim S_{11}^{-0.97}, \phi(\mbox{B}) \sim S_{11}^{-2.59}$
so that
$\phi(\mbox{B}) \propto \phi(\mbox{Be})^{2.67}$.
Those exponents were
obtained by Bahcall and Ulrich from 1000 SSMs with input parameters
randomly distributed near the most probable values \cite{Bahcall-Ulrich}.

The nonstandard solar models (the low opacity models, the low Z model, and the
models with large $S_{11}$) illustrated in Fig.~\ref{fig_comb},
include physically unreasonable models with $S_{11}$ as large as 2.5 times
and $T_C$ as small as 0.97 times their most probable values. Within
their theoretical uncertainties, all model predict
$\phi(\mbox{B}) \propto \phi(\mbox{Be})^n$
with $n = 2.25 - 2.67$.  Most extremely nonstandard solar  models
 still lead to $^7$Be and $^8$B neutrino fluxes that are adequately
parameterized by
simple power laws, {\it e.g.,} in the central temperature or $S_{11}$.
(Exceptions include the maximum rate model \cite{Bahcall-Pinsonneault},
the WIMP model, and the model with $S_{34} = 0$ \cite{Bahcall-Ulrich}.) This
happens because, although the
Sun as whole is not self-homologous (polytropic), over the range of
temperatures and densities in the Sun's inner core
($r < 0.2 R_\odot$), 91\% of the neutrino and energy production derives
from the single $pp$ reaction, and all the nuclear reactions and opacities
in the present Sun can be approximated by power laws.  Consequently, when
the luminosity is held constant, large changes in input parameters lead only
to nearly homologous changes in core temperatures, mass, and radius.

So far, we have shown that the
Kamiokande and Homestake results together, if correct,  essentially exclude
any astrophysical solutions.  What if either experiment were
wrong?  In Fig.~\ref{fig_cl}, we show the enlarged allowed region of the
combined fit when Homestake's quoted experimental error is tripled.  The data
still strongly disfavor the nonstandard
solar models: the cooler sun with $T_C$ reduced by 5\% is only allowed at
$\sim$1\% C.L.  Expressed otherwise, the best fit with
$\phi(\mbox{Be}) = 0$ corresponds to a Homestake rate of 2.9 SNU;
the best cool sun fit when the Homestake error is tripled is 3.3
SNU, compared with the value of $2.23 \pm 0.23$ in Table~I.
We have also carried out a calculation with
the cross section uncertainties for the chlorine and the
gallium detectors  increased by factors of three, and obtained a similar
result (Fig.~\ref{fig_xsec}).
Of course, if one entirely disregards either of these two experiments,
a large class of nonstandard models become possible.

In summary we have considered the most general solar model with
minimal constraints using the neutrino fluxes as free parameters, and shown
that the fit is excluded by the solar neutrino data at 98\% C.L., {\it i.e.},
essentially any astrophysical solution is incompatible with the
quoted data.  Furthermore,  this very improbable best fit point requires
$\phi(\mbox{Be})/\phi(\mbox{Be})_{SSM} < 0.07$
and
$\phi(\mbox{B})/\phi(\mbox{B})_{SSM} = 0.37 \pm 0.04$ (1$\sigma$),
which is inconsistent with virtually all explicit nonstandard solar models,
which predict a larger reduction of the $^8$B flux than the $^7$Be flux.
Increasing  the Homestake experimental error or the detector cross section
errors by factors three does not justify the nonstandard solar models.

We conclude that at least one of our original assumptions are wrong,
either (1) Some mechanism other than the $pp$ and CNO chains generates the
solar luminosity, or the Sun is not
in quasi-static equilibrium; (2) The neutrino energy spectrum is distorted
by some mechanism such as the MSW effect; (3) Either the Kamiokande or
Homestake result is grossly wrong.

We also noted that almost all explicit
nonstandard models fall on a narrow band in the
$\phi(\mbox{Be})-\phi(\mbox{B})$
plane, and can be characterized by a single effective parameter, the core
temperature \cite{BKL,BHKL}.

It is pleasure to thank Eugene Beier for useful discussions.
This work is supported by the Department of Energy Contract
DE-AC02-76-ERO-3071.

\newpage



\newpage

\begin{table}[p]
\caption{
%
%
The standard solar model predictions of Bahcall and Pinsonneault
\protect\cite{Bahcall-Pinsonneault} and of
Turck-Chi\'eze and Lopes
\protect\cite{Turck-Chieze-Lopes}, along with the results of the solar
neutrino experiments.
}
\label{}
\vspace{1.0ex}
\begin{tabular}{l  c c c}
%
               & BP SSM        & TCL SSM      & Experiments \\
\hline
Kamiokande     &  1 $\pm$ 0.14 & 0.77$\pm$0.19 & 0.50$\pm$0.07 BP-SSM  \\
Homestake (Cl) &  8$\pm$1 SNU  & 6.4$\pm$1.4 SNU & 2.23$\pm$0.23 SNU
                                                  (0.28$\pm$0.03 BP-SSM) \\
SAGE \& GALLEX (Ga) & 131.5$^{+7}_{-6}$ SNU & 122.5$\pm$7 SNU & 71$\pm$15 SNU
                                                      (0.54$\pm$0.11 BP-SSM) \\
%
\end{tabular}
\end{table}

\newpage


\begin{figure}[p]
\vspace{2ex}
\caption{
%
%
Each experiment is fit to the $pp$, $^7$Be, $^8$B, and CNO fluxes, imposing
only the luminosity constraint.  The fit neutrino fluxes are plotted in the
$^7$Be-$^8$B plane in units of the Bahcall-Pinsonneault predicted fluxes
\protect\cite{Bahcall-Pinsonneault}.
This parameter space represents all possible astrophysical solutions consistent
with our (minimal) assumptions.
The 90\% C.L.     uncertainties of the  Bahcall-Pinsonneault SSM are shown
in the upper-right corner.
}
\label{fig_exps}
\end{figure}

\begin{figure}[p]
\vspace{2ex}
\caption{
%
%
The allowed region from the  combined fit of the Kamiokande, Homestake, and
gallium results at 90, 95, and 99\% C.L. allowing (unphysical) negative
values for $\phi$(Be).  For $\phi(\mbox{Be}) \geq 0$,  $\chi^2 = 5.6$ for
the best fit, and the model is excluded at 98\% C.L. for 1 degree of freedom.
Therefore any astrophysical solution is excluded at $\geq 98\%$ C.L. Also shown
are various nonstandard solar models and the power laws of the central
temperature ($T_C$) and the $pp$ cross section ($S_{11}$) obtained from the
Bahcall-Ulrich SSMs \protect\cite{Bahcall-Ulrich}, and extrapolated from the
SSM region.  All nonstandard models other than WIMPs can be approximately
parameterized by $T_C$ or $S_{11}$.
The error bars show the uncertainty of $\phi$(B) due to the $p$+$^7$Be cross
section \protect\cite{Bahcall-Pinsonneault}.
}
\label{fig_comb}
\end{figure}

\begin{figure}[p]
\vspace{2ex}
\caption{
%
%
The combined fit when the Homestake experimental error is tripled.
}
\label{fig_cl}
\end{figure}

\begin{figure}[p]
\vspace{2ex}
\caption{
%
%
The combined fit when the detector cross section uncertainties are tripled
from 3.3\% (Homestake) and 4\% (gallium).
}
\label{fig_xsec}
\end{figure}

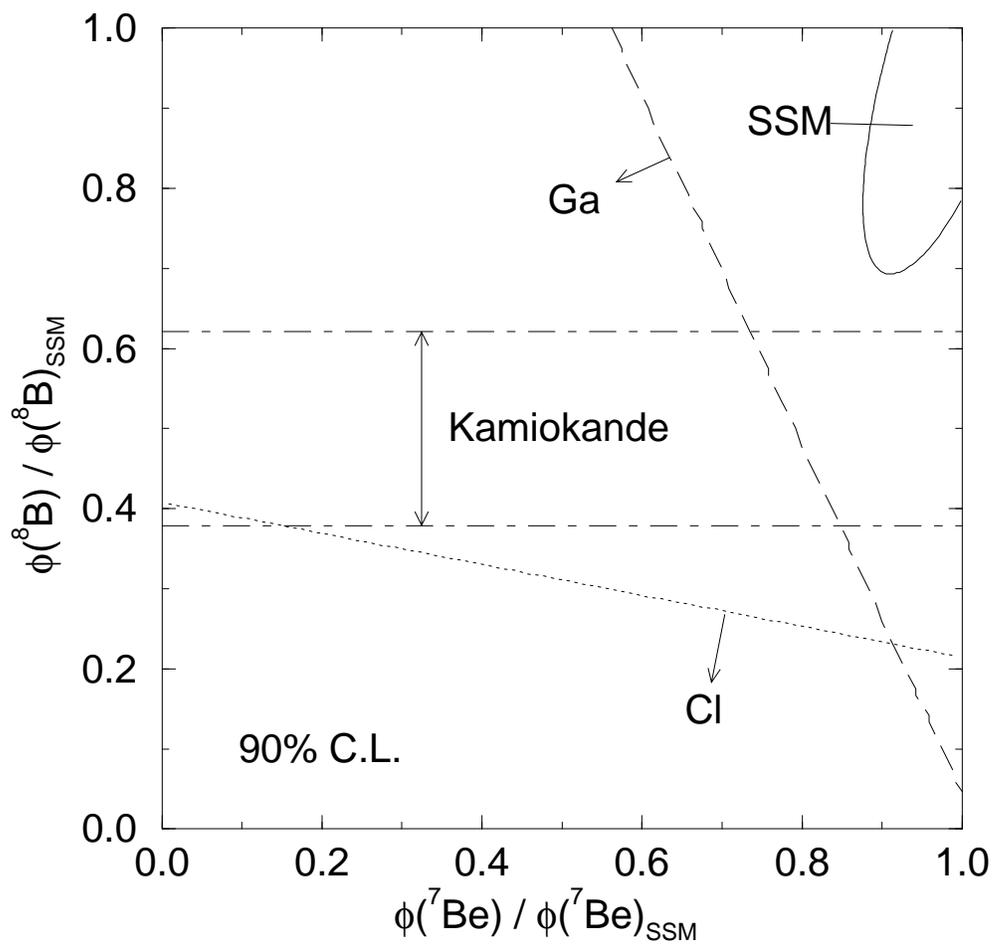
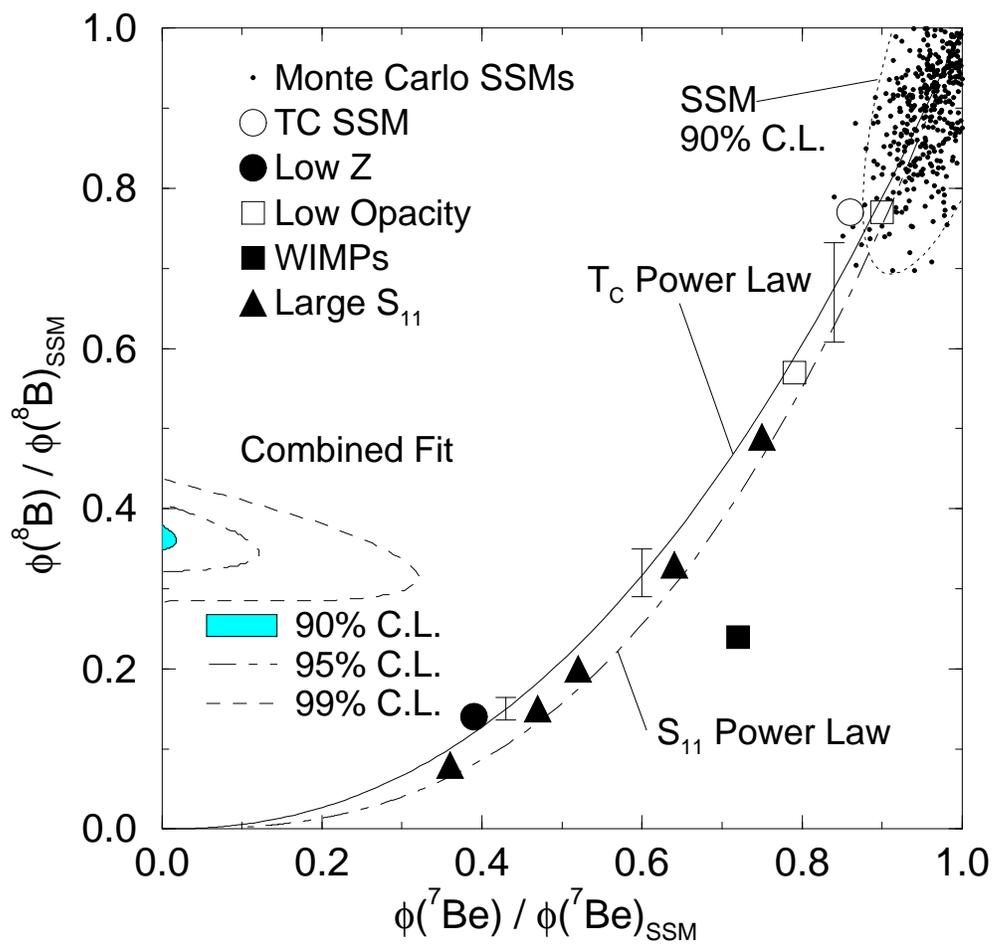
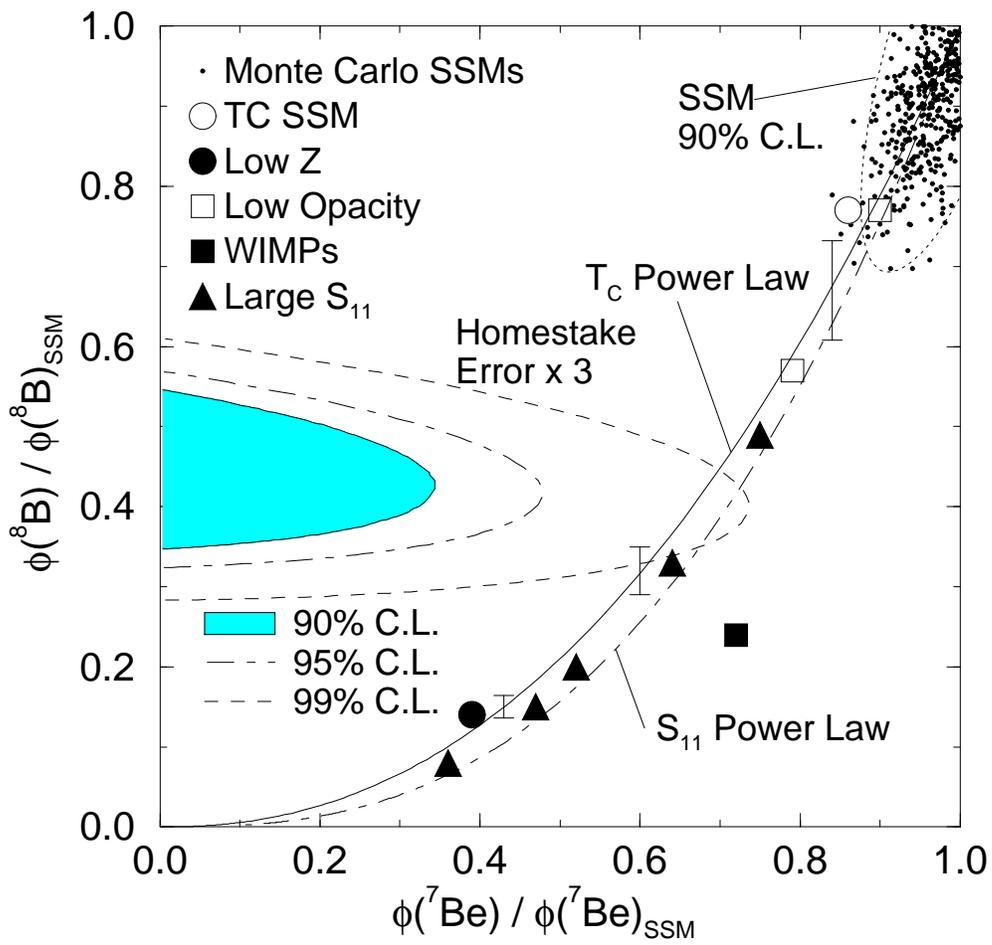
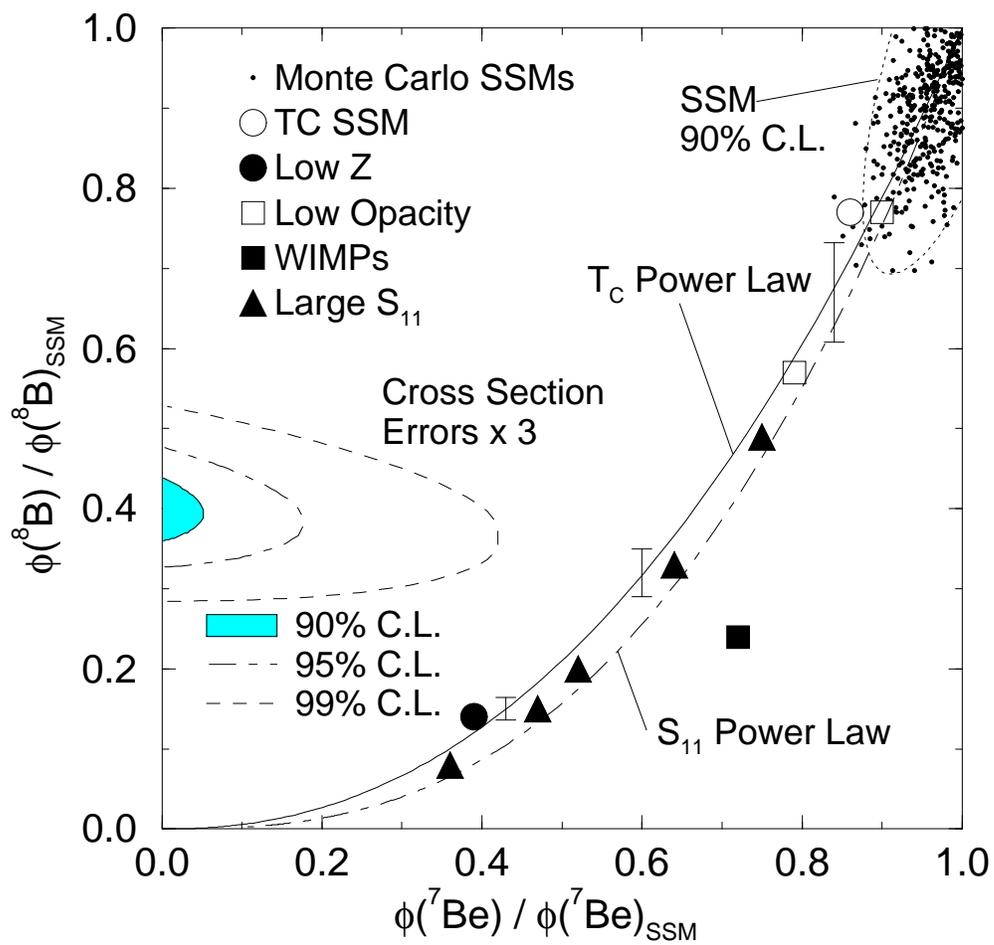
\end{document}